\documentstyle{llncs}

\begin{document}

\title{Generalized Grover Search Algorithm for Arbitrary Initial
Amplitude Distribution}

\author{David Biron,$^1$  Ofer Biham,$^1$  Eli Biham,$^2$ Markus
Grassl,$^3$  Daniel A. Lidar$^4$}

\institute{
$^1$Racah Institute of Physics, The Hebrew University, Jerusalem
91904, Israel\\ 
$^2$Computer Science Department, Technion, Haifa 32000, Israel \\
$^3$Institut f\"{u}r Algorithmen und Kognitive Systeme,
Universit\"{a}t Karlsruhe, 
\hbox{Am Fasanengarten 5}, D--76128 Karlsruhe, Germany\\
$^4$Department of Chemistry, University of California, Berkeley, CA
94720, USA}

\maketitle

\begin{abstract}
Grover's algorithm for quantum searching of a database is generalized
to deal with arbitrary initial amplitude distributions. First order
linear difference equations are found for the time evolution of the
amplitudes of the $r$ marked and $N\!-\!r$ unmarked states. These
equations are solved {\it exactly}. An expression for the optimal
measurement time $T \sim O(\sqrt{N/r})$ is derived which is shown to
depend only on the initial average amplitudes of the marked and
unmarked states. A bound on the probability of measuring a marked
state is derived, which depends only on the standard deviation of the
initial amplitude distributions of the marked or unmarked states.

\end{abstract}

\section{Introduction}

\label{intro}

The power of Quantum Computation (QC) was most dramatically
demonstrated in the algorithms of Shor, for the polynomial time
solution of the factorization problem \cite{Shor:94}, and of Grover
\cite{Grover96,Grover97}, for a search which can find a marked element
in an unsorted database of size $N$, in $O(\sqrt{N})$ steps (compared
to $O(N)$ steps on a classical computer). The importance of Grover's
result stems from the fact that it proves the enhanced power of
quantum computers compared to classical ones for a whole class of
problems, for which the bound on the efficiency of classical
algorithms is known. This is unlike the case of Shor's algorithm,
since in spite of the fact that no efficient classical algorithm for
the factorization problem is known, there is no proof that such an
algorithm does not exist.

A large number of related results followed Grover's original paper
\cite {Grover96}. Among these, the efficiency of Grover's algorithm
was analyzed and compared to the theoretical efficiency limit of
quantum computers for such benchmark search problems as introduced
(before Grover's result \cite{Grover96})
by {Bennett et al.} \cite{Bennett97}. The algorithm was recently shown
to be optimal, i.e., to satisfy the theoretical limit
\cite{Zalka:97}. Further developments include the use of Grover's
algorithm or slightly modified versions of it as the essential step in
algorithms that solve a variety of other problems such as quantum
search for the median \cite{Grover97b} and the minimum \cite{Durr96}
in a set of $N$ items, as well as the collision problem
\cite{Brassard97a}. It was also shown that other search problems which
classically require $\log _{2}N$ evaluations ({\em queries}) of a
black--box function, can be reduced to a single query using Grover's
approach \cite{Grover97c,Terhal97}. Finally, it was shown that a
simple closed formula describes the time evolution of the amplitudes
of the generalized problem, which includes several marked states
\cite{Boyer96}. As this work is directly relevant to ours, we briefly
summarize some of its pertinent results.

Let $k(t)$ [$l(t)$] denote the amplitude of the {\it marked} [{\it
unmarked\/}] states in the database, $r$ the number of marked states,
and $\omega =2\arcsin (\sqrt{r/N})$. It was shown by Boyer et
al. \cite{Boyer96} that after $t$ steps of the algorithm, the marked
states' amplitude increases as: $k(t)=\sin [\omega (t+1/2)]/\sqrt{r}$,
while at the same time that of the unmarked states decreases as:
$l(t)=\cos [\omega (t+1/2)]/\sqrt{ N-r}$. Since $N$ is large, the
optimal time to measure and complete the calculation is thus after
$T=O(\sqrt{N/r})$ time steps, when $k(t)$ is maximal. This analysis
relies on the fact that the initial amplitude distribution is {\it
uniform}. However, in a variety of interesting cases it would be
desirable to apply Grover's algorithm to a {\it non}-uniform initial
distribution. Generically, this could arise in situations where the
search is used as a subroutine in a larger quantum computation, and
the input to the algorithm can thus not be controlled. Another example
would be cases where the given initial distribution over the
elements is intrinsically non-uniform.

In this paper we generalize Grover's algorithm to the case in which
the initial amplitudes are either real or complex and follow {\it any
arbitrary distribution}. The time evolution of the amplitudes is
solved {\it exactly} for general initial conditions, and the
efficiency of the algorithm is evaluated. It is found that for generic
initial conditions, the search algorithm still requires
$O(\sqrt{N/r})$ steps, with only a constant factor compared to the
case of a uniform initial distribution \cite{Grover97}.

The paper is organized as follows. In Sec. \ref{rec-eq} we define the
modified Grover algorithm and derive difference equations for the time
evolution of the amplitudes in it. We solve these equations exactly in
Sec.  \ref{solution}, and analyze the results in
Sec. \ref{analysis}. A summary and conclusions are presented in
Sec. \ref{conclusions}.

\section{The Recursion Equations}

\label{rec-eq}

\subsection{The Generalized Algorithm}

Our modified algorithm is essentially Grover's original algorithm, but
without the initialization step. It thus consists of the following
stages: 

\begin{enumerate}
\item Use any initial distribution of marked and unmarked states,
e.g., the final state of any other quantum algorithm (do {\it not}
initialize the system to the uniform distribution).

\item  Repeat the following steps $T$ times [an expression for $T$ is
given in Eq.~(\ref{eq:th-T})]:

\begin{description}
\item[A.]  Rotate the marked states by a phase of $\pi $ radians.

\item[B.]  Rotate all states by $\pi $ radians around the average
amplitude of {\it all} states. This is done by applying the 
``inversion about average'' operator, represented by the following
unitary matrix:

\[
D_{i,j}=\left\{ 
\begin{array}{ll}
\frac{2}{N} & \mbox{ if }i\neq j \\ 
\noalign{\medskip}
\frac{2}{N}-1 & \mbox{ if }i=j
\end{array}
\right. 
\]
\end{description}

\item  Measure the resulting state.
\end{enumerate}

\subsection{The Dynamics}

We will now analyze the time evolution of the amplitudes in the
modified algorithm with a total of $N$ states. Let the marked
amplitudes at time $t$ be denoted by $k_{i}(t)$, $i=1,\dots ,r$ and
the unmarked amplitudes by $ l_{i}(t)$, $i=r+1,\dots ,N$, where the
initial distribution at $t=0$ is arbitrary. Without loss of generality
we assume that the number of marked states satisfies $1\leq r\leq
N/2$. We denote the averages and variances of the amplitudes by:
\begin{eqnarray}
\mbox{marked:} &&  
\rlap{$\displaystyle\bar{k}(t)=\frac{1}{r} \sum_{j=1}^{r}k_{j}(t)$}
\phantom{\bar{l}(t)=\frac{1}{N-r} \sum_{j=r+1}^{N}l_{j}(t)}\qquad
 \sigma _{k}^{2}(t)=\frac{1}{r} \sum_{j=1}^{r}|k_{j}(t)-\bar{k}(t)|^{2}
\label{eq:marked} \\
\mbox{unmarked:} && \bar{l}(t)=\frac{1}{N-r} \sum_{j=r+1}^{N}l_{j}(t)
\qquad \sigma _{l}^{2}(t)=\frac{1}{N-r}
\sum_{j=r+1}^{N}|l_{j}(t)-\bar{l} (t)|^{2}\quad  \label{eq:unmarked}
\end{eqnarray}
The key observation is that the entire dynamics dictated by Grover's
algorithm can be described in full by the time-dependence of the {\it 
averages}.
(The variances are defined above for convenience, as they
are used later in a different context -- see Section~\ref{max}.)
Formally, let:  
\begin{equation}
C(t)=-\frac{2}{N}\left[
\sum_{j=1}^{r}k_{j}(t)-\sum_{j=r+1}^{N}l_{j}(t)\right]
=\frac{2}{N}\left[ 
(N-r)\bar{l}(t)-r\bar{k}(t)\right] {\rm \thinspace .}  \label{eq:C}
\end{equation}
$C(t)$ is thus the weighted average over the marked and unmarked
states, with the minus sign accounting for the $\pi $ phase difference
between them during the algorithm iterations. The following theorem
then shows that all states evolve equally:

\begin{theorem}
The time evolution of all amplitudes (of both marked and unmarked
states) is 
independent of the state index, and satisfies: 
\begin{eqnarray}
k_{i}(t+1) &=&C(t)+k_{i}(t)\qquad\qquad i=1,\dots ,r
\label{eq:recur_k} \\
l_{i}(t+1)
&=&\rlap{$C(t)-l_{i}(t)$}\phantom{C(t)+k_{i}(t)}\qquad\qquad
i=r+1,\dots ,N 
\label{eq:recur_l}
\end{eqnarray}
\end{theorem}
{\it Proof. --}
This follows directly from the algorithm. Consider any marked state
$k_{i}(t)$; this state is flipped to $k_{i}^{\prime }(t)=-k_{i}(t)$,
so that the marked average becomes $\bar{k}^{\prime }(t)=\frac{1}{r}
\sum_{j=1}^{r}k_{j}^{\prime }(t)=-\bar{k}(t)$. The unmarked states, on
the other hand, do not flip, so that the total average after the flip
is: $x(t)= \frac{1}{N}[r\,\bar{k}^{\prime
}(t)+(N-r)\,\bar{l}(t)]=C(t)/2$. ``Inversion about average'' is by
definition: $k_{i}^{\prime }(t)\rightarrow 2x(t)-k_{i}^{\prime }(t)$
and $l_{i}(t)\rightarrow 2x(t)-l_{i}(t)$.  Therefore in total:\
$k_{i}(t)\rightarrow C(t)+k_{i}(t)$ and $ l_{i}(t)\rightarrow
C(t)-l_{i}(t)$. \qed

\noindent From this it follows by averaging that:

\begin{corollary}
The average marked and unmarked amplitudes obey first order linear
coupled difference equations:
\end{corollary}

\begin{eqnarray}
\bar{k}(t+1) &=&C(t)+\bar{k}(t)  \label{eq:k-ave} \\
\bar{l}(t+1) &=&C(t)-\bar{l}(t){\rm .}  \label{eq:l-ave}
\end{eqnarray}
These equations can be solved for $\bar{k}(t)$ and $\bar{l}(t)$, and
along with the initial distribution this yield the exact solution for
the dynamics of all amplitudes by using Eqs.~(\ref{eq:recur_k}) and
(\ref{eq:recur_l}).
\section{Solution of the Recursion Equations}

\label{solution}

The recursion formulae \noindent can be solved by a standard
diagonalization method for arbitrary complex initial conditions. Let:

\[
{\bf v}(t)=\left( \bar{k}(t),\bar{l}(t)\right) {\rm ,} 
\]

\noindent and define:

\[
a\equiv \frac{N-2r}{N}\,{\rm ,}\quad \quad b\equiv \frac{2(N-r)}{N}\,
{\rm , }\quad \quad c\equiv \frac{2r}{N}{\rm .} 
\]

\noindent The recursion equations (\ref{eq:k-ave}) and
(\ref{eq:l-ave}) can be written as: 
\[
{\bf v}(t+1)={\sf A}\cdot {\bf v}(t)\,{\rm ,}\qquad {\sf A}
=\left( 
\begin{array}{rr}
a & b \\ 
-c & a
\end{array}
\right) {\rm \thinspace .} 
\]
Diagonalization of ${\sf A}$ yields a solution for ${\bf v}(t)$, as
follows. Let ${\sf S}$ be the diagonalizing matrix: 
\[
{\sf A}^{D}\equiv {\sf S}^{-1}{\sf AS=}\left( 
\begin{array}{ll}
\lambda _{-} & 0 \\ 
0 & \lambda _{+}
\end{array}
\right) {\rm ,}\quad \quad \lambda _{\pm }=\gamma \,e^{\pm i\omega }
{\rm .} 
\]
Then ${\bf w}(t)={\sf S}^{-1}\cdot {\bf v}(t)$ satisfies: 
\[
{\bf w}(t+1)={\sf A}^{D}\cdot {\bf w}(t)\,{\rm ,} 
\]
with solution: 
\[
{\bf w}(t)=\left( (\lambda _{-})^{t}\,w_{-}(0),(\lambda
_{+})^{t}\,w_{+}(0)\right)
\]
where ${\bf w}(0)=(w_{-}(0),w_{+}(0))$.
This yields $\bar{k}(t)$ and $\bar{l}(t)$ from ${\bf v}(t)={\sf S}
\cdot {\bf w}(t)$. Diagonalizing ${\sf A}$ one finds: 
\begin{eqnarray}
\gamma &=&a^{2}+bc=1  \label{eq:alpha} \\
\cos \omega &=&a=1-2\frac{r}{N}{\rm ,}  \label{eq:omega}
\end{eqnarray}
which is identical to the frequency found by Boyer et
al. \cite{Boyer96} The 
eigenvectors of ${\sf A}$ are the columns of ${\sf S}$: 
\[
{\sf S}=\left( 
\begin{array}{c@{\quad}c}
i\sqrt{\frac{N}{r}-1} & -i\sqrt{\frac{N}{r}-1} \\ 
1 & 1
\end{array}
\right) \,{\rm ,\quad }{\sf S}^{-1}=\left( 
\begin{array}{r@{\quad}r}
-\frac{i}{2}\sqrt{\frac{r}{N-r}} & \frac{1}{2} \\ 
\frac{i}{2}\sqrt{\frac{r}{N-r}} & \frac{1}{2}
\end{array}
\right) {\rm .} 
\]
Using this: 
\[
\left( 
\begin{array}{l}
w_{-}(0) \\ 
w_{+}(0)
\end{array}
\right) ={\bf w(}0)={\sf S}^{-1}\cdot {\bf v}(0)=\left( 
\begin{array}{r}
-\frac{i}{2}\sqrt{\frac{r}{N-r}}\bar{k}(0)+\frac{1}{2}\bar{l}(0) \\ 
\frac{i}{2}\sqrt{\frac{r}{N-r}}\bar{k}(0)+\frac{1}{2}\bar{l}(0)
\end{array}
\right) {\rm ,} 
\]
so that: 
\[
{\bf v}(t)={\sf S}\cdot \left( 
\begin{array}{c}
\left( -\frac{i}{2}\sqrt{\frac{r}{N-r}}\bar{k}(0)+\frac{1}{2}\bar{l}
(0)\right) e^{-i\omega t} \\ 
\left( \frac{i}{2}\sqrt{\frac{r}{N-r}}\bar{k}(0)+\frac{1}{2}\bar{l}
(0)\right) e^{i\omega t}
\end{array}
\right) {\rm .} 
\]
This yields finally, after some straightforward algebra: 
\begin{eqnarray}
\bar{k}(t) &=&\bar{k}(0)\cos \omega
t+\bar{l}(0)\sqrt{\frac{N-r}{r}}\,\sin 
\omega t  \label{eq:k(t)} \\
\bar{l}(t) &=&\bar{l}(0)\cos \omega t-\bar{k}(0)\sqrt{\frac{r}{N-r}}\sin
\omega t{\rm .}  \label{eq:l(t)}
\end{eqnarray}
Together with Eqs. (\ref{eq:recur_k}) and (\ref{eq:recur_l}) this
provides the complete exact solution to the dynamics of the amplitudes
in the generalized Grover algorithm, for arbitrary initial conditions.

\section{Analysis}

\label{analysis}

Next we derive several properties of the amplitudes.

\subsection{Phase Difference}

The averaged amplitudes can be expressed concisely as follows (even
when $ \bar{k}(0)$ and $\bar{l}(0)$ are complex):

\begin{eqnarray}
\bar{k}(t) &=&\alpha \sin (\omega t+\phi )  \label{eq:ksin} \\
\bar{l}(t) &=&\beta \cos (\omega t+\phi )  \label{eq:lcos}
\end{eqnarray}
where

\begin{eqnarray}
\tan \phi = \frac{\bar{k}(0)}{\bar{l}(0)}\sqrt{\frac{r}{N-r}}\,;
\qquad
\alpha ^{2}&=&
\bar{k}(0)^{2}+\bar{l}(0)^{2}\frac{N-r}{r}\,;\nonumber
\\\beta^{2}&=&
\bar{l}(0)^{2}+\bar{k}(0)^{2}\frac{r}{N-r}  \label{eq:defs}
\end{eqnarray}

\noindent which shows that there is a $\pi /2$ phase difference between
the marked and unmarked amplitudes: when the average marked amplitude
is maximal, the average unmarked amplitude is minimal, and {\it vice
versa}.

\subsection{Constant Variance}

Subtracting Eq.~(\ref{eq:recur_k}) from Eq.~(\ref{eq:k-ave}), and
subtracting Eq.~(\ref{eq:recur_l}) from Eq.~(\ref{eq:l-ave}), one
finds:

\begin{eqnarray}
k_i(t+1)-\bar{k}(t+1) &=& k_i(t)-\bar{k}(t)  \label{eq:kdiff} \\
l_i(t+1)-\bar{l}(t+1) &=& -[l_i(t)-\bar{l}(t)] {\rm .}  \label{eq:ldiff}
\end{eqnarray}

\noindent This means that:

\begin{eqnarray}
\Delta k_i \equiv k_i(t)-\bar{k}(t) \quad {\rm and} \quad \Delta l_i
\equiv (-1)^t [l_i(t)-\bar{l}(t)] {\rm ,}
\end{eqnarray}

\noindent are {\it constants of motion} (time-independent). It follows
immediately from the definition that the variances $\sigma _{k}^{2}$
and $\sigma _{l}^{2}$ 
[cf.~Eqs.~(\ref{eq:marked}) and (\ref{eq:unmarked})]
too, are both time-independent.

This allows us to simplify the expression for the time dependence of the
amplitudes:

\begin{eqnarray}
k_i(t) &=& \bar{k}(t) + \Delta k_i \\
l_i(t) &=& \bar{l}(t) + (-1)^t \Delta l_i {\rm ,}
\end{eqnarray}

\noindent where $\Delta k_i$ and $\Delta l_i$ are evaluated at $t=0$.

\subsection{Maximal Probability of Success and Optimal Number of
Iterations}
\label{max}

The probability that a marked state will be obtained in the
measurement at time $t$ at the end of the process is
$P(t)=\sum_{i=1}^{r}|k_{i}(t)|^{2}$. A bound on this quantity can be
derived as follows. Since all the operators used are unitary, the
amplitudes satisfy the normalization condition:

\begin{equation}
\sum_{i=1}^{r}|k_{i}(t)|^{2}+\sum_{i=r+1}^{N}|l_{i}(t)|^{2}=1
\label{eq:norm}
\end{equation}
at all times. Using
$\overline{(y-\bar{y})^{2}}=\overline{y^{2}}-\bar{y}^{2}$ 
($y$ is a random variable), we find from Eq.~(\ref{eq:unmarked}):
$$
\sum_{i=r+1}^{N}|l_{i}(t)|^{2}=(N-r)\sigma
_{l}^{2}+|\sum_{i=r+1}^{N}l_{i}(t)|^{2}/(N-r).
$$ Let:

\begin{equation} P_{\max }=1-(N-r)\sigma _{l}^{2}{\rm ,}
\label{eq:bound}
\end{equation}
a time-{\it independent }quantity. Note that in the case of a uniform
initial distribution of amplitudes $\sigma _{l}^{2}=0$ and $P_{\max
}=1$. Now, $ P(t)=P_{\max }-(N-r)|\bar{l}(t)|^{2}$, so that:

\begin{equation} P(t)\leq P_{\max }
\end{equation}
is the required bound. Using the exact solution, we can show that the
$ P_{\max }$ bound is in fact tight. For, from Eq.~(\ref{eq:lcos}) it
follows that $\bar{l}(T)=0$ when:

\begin{equation} \omega T+\phi =(j+1/2)\pi \,,\quad j=0,1,2,...
\label{eq:T}
\end{equation}
At these times the bound is reached so that times $T$ satisfying
Eq.~(\ref {eq:T}) are optimal for measurement. Note that this
conclusion holds only if $\bar{k}(0)/\bar{l}(0)$ is {\em real}. When
$\bar{k}(0)/\bar{l}(0)$ is complex, the bound is generally not reached
since $\bar{l}(t)$ may never vanish.  Collecting our results:

\begin{theorem}
Given arbitrary initial distributions of $r$ marked and $N\!-\!r$
unmarked states, with known averages $\bar{k}(0)$ and $\bar{l}(0)$
respectively, $ \bar{k}(0)/\bar{l}(0)$ real, the optimal measurement
times are after:

\begin{equation}
T={\frac{{(j+1/2)\pi -\arctan \left[ \
\frac{\bar{k}(0)}{\bar{l}(0)}\sqrt{ 
\frac{r}{N-r}}\right] }}{\arccos \left( 1-2\frac{r}{N}\right) }\,,}\quad
j=0,1,2,...  \label{eq:th-T}
\end{equation}
steps, when the probability of obtaining a marked state is $P_{\max}$
as given by Eq.~(\ref{eq:bound}).
\end{theorem}

An important conclusion is that to determine the optimal measurement
times, all one needs to know are the average initial amplitudes and
the number of marked states. The more difficult case when these are
unavailable will be considered in a separate publication \cite{TBP}.
The expansion of
Eq.~(\ref{eq:th-T}) in $r/N\ll 1$ (at $j=0$) yields: 
\begin{equation}
T = -\frac{1}{2}\frac{\bar{k}(0)}{\bar{l}(0)} + 
\frac{\pi}{4} \sqrt{N/r} - \frac{\pi}{24} \sqrt{r/N} + O(r/N)
{\rm ,} 
\label{eq:Tapprox}
\end{equation}
confirming that Grover's algorithm converges in $O(\sqrt{N/r})$ steps
for arbitrary distributions. The advantage of an initial amplitude
distribution with a relatively high average of the marked states is
manifested in 
the constant offset $-\frac{1}{2}\frac{\bar{k}(0)}{\bar{l}(0)}$, 
which may significantly reduce the required
number of steps.

\section{Summary and Conclusions}

\label{conclusions}
In this work we generalized Grover's quantum search algorithm to apply
for initial input distributions which are non-uniform. In fact, it was
shown that by simply omitting the first step of Grover's original
algorithm, wherein a uniform superposition is created over all
elements in the database, a more general algorithm results which
applies to {\it arbitrary } initial distributions. To analyze the
algorithm, we found that the time evolution of the amplitudes of the
marked and unmarked states can be described by first-order linear
difference equations with some special properties. The most important
of these is that all amplitudes essentially evolve uniformly, with the
dynamics being determined completely by the average amplitudes. This
observation allowed us to find an exact solution for the
time-evolution of the amplitudes. An important conclusion from this
solution is that generically the generalized algorithm also has a
$O(\sqrt{ N/r})$ running time, thus being more powerful than any
classical algorithm designed to solve the same task.

This work was initiated during the 1997 Elsag-Bailey --
I.S.I. Foundation research meeting on quantum computation.


\begin{thebibliography}{10}

\bibitem{Shor:94} {P.W. Shor}, {Polynomial-time algorithms for prime
factorization and discrete logarithms 
     on a quantum computer}, SIAM Journal on Computing, {\bf 26},
1484 (1997).

\bibitem{Grover96}  {L. Grover}, 
{A fast quantum mechanical algorithm for database search},
in {\it Proceedings of the
Twenty-Eighth 
Annual Symposium on the Theory of Computing}, {ACM Press} ({New York},
1996), p.\ 212.

\bibitem{Grover97}  {L. Grover}, 
{Quantum mechanics helps in searching for a needle in a haystack},
Phys. Rev. Lett. {\bf 79}, 325 (1997).

\bibitem{Bennett97}  {C.H. Bennett, E. Bernstein, G. Brassard, and U.
Vazirani}, 
{Strengths and weaknesses of quantum computing}, SIAM
Journal on Computing {\bf 26}, 1510 (1997).

\bibitem{Zalka:97}  {C. Zalka}, {Grover's quantum searching algorithm is
optimal (\uppercase{L}ANL preprint quant-ph/9711070)}.

\bibitem{Grover97b}  {L. Grover}, {Quantum telecomputation
(\uppercase{L}ANL preprint quant-ph/9704012)}.

\bibitem{Durr96}  {C. Durr and P. Hoyer}, {A quantum algorithm for
finding the minimum (\uppercase{L}ANL preprint quant-ph/9607014)}.

\bibitem{Brassard97a}  {G. Brassard, P. Hoyer, and A. Tapp}, {Quantum
algorithm for the collision problem (\uppercase{L}ANL preprint
quant-ph/9705002)}.

\bibitem{Grover97c}  {L. Grover}, {Quantum computers can search
arbitrarily large databases by a single query}, Phys. Rev. Lett. {\bf
79}, 4709 (1997).

\bibitem{Terhal97}  {B.M. Terhal and J.A. Smolin}, {Single quantum
querying of a database (\uppercase{L}ANL preprint quant-ph/9705041)}.

\bibitem{Boyer96} {M. Boyer, G. Brassard, P. Hoyer and A. Tapp}, 
{Tight bounds on quantum searching},
in
{\it \ Proceedings of the fourth workshop on Physics and Computation},
edited by {\ T. Toffoli, M. Biafore and J. Leao, New England Complex
Systems Institute}, ({Boston}, 1996), p.\ 36. To appear in
Fortschritte der Physik.

\bibitem{TBP} D. Biron, O. Biham, E. Biham, M. Grassl, and D.A. Lidar,
to be published.

\end{thebibliography}
\end{document}